\documentclass[twocolumn]{aastex701} 

\usepackage{multirow}

\usepackage{graphics,epsf}
\usepackage[utf8]{inputenc}
\usepackage{amsmath}                % American Mathematical Society package
\usepackage{amsfonts}               % American Mathematical Society fonts
\usepackage{amssymb}                % American Mathematical Society symbol
\usepackage{epsfig}                 % EPS figures
\usepackage{graphicx}               % Required for inserting images
\usepackage{float}
\usepackage{color}
\usepackage{multirow}               % double row table entries

\hypersetup{
    colorlinks=true,
    linkcolor=red,   
    urlcolor=cyan}

\usepackage[colorinlistoftodos]{todonotes}

\newcommand{\cm}{{~\rm cm}}
\newcommand{\km}{{~\rm km}}
\newcommand{\s}{{~\rm s}}

\newcommand{\g}{{~\rm g}}
\newcommand{\G}{{~\rm G}}
\newcommand{\K}{{~\rm K}}
\newcommand{\erg}{{~\rm erg}}
\newcommand{\yr}{{~\rm yr}}

\newcommand{\keV}{{~\rm keV}}
\newcommand{\kev}{{~\rm keV}}

\begin{document}

\title{Reproducing morphological features in the  supernova remnant G11.2-0.3 by simulating jittering jets}

%\title{Simulating the jittering-jets explosion mechanism: Supernova remnant G11.2-0.3}

%\author{Muhammad Akashi\,\orcidlink{0000-0001-7233-6871}}
%\affiliation{Kinneret College on the Sea of Galilee, Samakh 15132, Israel}
%\affiliation{Department of Physics, Technion - Israel Institute of Technology, Haifa, 3200003, Israel; akashi@technion.ac.il}

%\author{Noam Soker\,\orcidlink{0000-0003-0375-8987}} 
%\affiliation{Department of Physics, Technion - Israel Institute of Technology, Haifa, 3200003, Israel; soker@technion.ac.il; }

\author[0000-0001-7233-6871]{Muhammad Akashi}
\affiliation{Kinneret College on the Sea of Galilee, Samakh 15132, Israel}
\affiliation{Department of Physics, Technion - Israel Institute of Technology, Haifa, 3200003, Israel; 
akashi@technion.ac.il; soker@physics.technion.ac.il}
\email{akashi@technion.ac.il}

\author[0000-0003-0375-8987]{Noam Soker}
\affiliation{Department of Physics, Technion - Israel Institute of Technology, Haifa, 3200003, Israel; 
akashi@technion.ac.il; soker@physics.technion.ac.il}
\email{soker@physics.technion.ac.il}

%\date{\today}

\begin{abstract}
We hydrodynamically simulate a core-collapse supernova (CCSN) explosion by launching three pairs of jets in the framework of the jittering-jets explosion mechanism (JJEM), and reproduce a morphology of two opposite circum-jet rings and a bar of dense gas perpendicular to the rings' axis, resembling these morphological features in the CCSN remnant SNR G11.2-0.3. 
The first pair of wide jets is very energetic; it triggers the explosion and inflates two bubbles that compress the material in an expanding shell. The bubbles also compress material in a plane perpendicular to the jet axis. The second pair of wide jets removes material from this plane, beside along a bar that is on an axis perpendicular to the two pairs' axes. The jets of the third pair, now of narrow jets, penetrate the expanding shell and compress material to their sides to form two opposite rings. These morphological features are qualitatively similar to those observed in the point-symmetric CCSN remnant G11.2-0.3. As competing theoretical CCSN explosion mechanisms cannot explain point-symmetric CCSN remnants, our study provides some support for the claim that the JJEM is the primary explosion mechanism of CCSNe. 
\end{abstract}
   

% ==================================
\section{Introduction} 
\label{sec:intro}
% ==================================

There are two intensively studied theoretical explosion mechanisms aiming to explain the majority of core-collapse supernova (CCSN) explosions: The neutrino-driven (delayed-neutrino; neutrino heating) explosion mechanism (e.g., \citealt{Akhmetalietal2026, ChenCHetal2026, EggenbergerAndersenetal2026, Giudicietal2026, LuoZhaKajino2026, Mezzacappa2026, Murphyetal2026, PanLi2026, Paradisoetal2026, Rusakovetal2026, VarmaMuller2026, Wessonetal2026} for some papers from 2026), and the jittering-jets explosion mechanism (JJEM; e.g., \citealt{Soker2026SN1987Amulecular, Soker2026SNRJ0450, Soker2026Failed}, for papers from 2026); for earlier reviews, see, e.g., \cite{Soker2025Learning} for the JJEM, and \cite{Janka2025} for the delayed-neutrino mechanism, and \cite{Soker2025G11} for the relation between them, and to other rare explosion mechanisms and other energy sources in CCSNe.  

In the JJEM, several to about twenty pairs of jets with fully or partially stochastic variations in their axes explode the star. 
Intermittent accretion disks around the newly born neutron star (NS) launch the jittering jets. The pre-collapse core rotation introduces a fixed angular momentum component to the accreted gas, while the pre-collapse core convection introduces the stochastic component (e.g., \citealt{Soker2023gap}). According to the assumption of the JJEM, vortices in the convection zones of the collapsing core seed instabilities above the NS (for some studies of these types of instabilities see, e.g., \citealt{Abdikamalovetal2016, KazeroniAbdikamalov2020, Buelletetal2023}) that amplify the stochastic angular momentum fluctuations (e.g., \citealt{ShishkinSoker2023, WangShishkinSoker2024, WangShishkinSoker2025}). 

The jets that explode the star in the framework of the JJEM might leave imprints on CCSN remnants (CCSNRs) in the form of jet-shaped morphological features, particularly point-symmetric morphologies. Point-symmetrical morphologies are misaligned pairs of opposite structural features (e.g., \citealt{ShishkinMichaelis2026}). The structural features might include rings, lobes, bubbles, clumps, filaments, nozzles, and ears, as three-dimensional (3D) hydrodynamic simulations of the JJEM have recently shown \citep{Braudoetal2025, SokerAkashi2025, AkashiSoker2026a, BraudoSoker2026}. 
In general, the jittering jets are not relativistic (\citealt{Guettaetal2020} argued that most CCSNe have no relativistic jets, unlike those of gamma ray bursts, e.g., \citealt{Izzoetal2019, AbdikamalovBeniamini2025}).
The morphologies of some CCSNRs exhibit one, two, or three very powerful pairs of jets, where in some cases the two jets in a pair differ substantially in their energy and momentum (e.g., \citealt{Bearetal2025Puppis, Shishkinetal2025S147}). Each energetic pair of jets can excite a strong shock in the exploding star, leading to a photospheric shell in the early stages of the explosion. Two or more energetic pairs can, therefore, form two or more photospheric shells that observations might reveal in the first days to weeks of the explosion. \cite{SokerShiran2025}  identified two (or possibly three) photospheric shells in SN 2023ixf (based on observation by \citealt{Zimmermanetal2024}), and \cite{ShiranSoker2026} identified two photospheric shells in SN 2024ggi (based on observations by \citealt{ChenTWetal2025}). 

At present, CCSNRs' morphologies are the only observable that can robustly distinguish between the two theoretical explosion mechanisms (e.g., \citealt{Soker2024UnivReview, Soker2025Learning}), and they clearly support the JJEM (e.g., \citealt{BearSoker2025, SokerShishkin2025Vela, ShishkinSoker2025Crab, Soker2025G0901, Soker2025G11, Klimovetal2026, KlimovSoker2026b}), and severely challenge the neutrino-driven mechanism, which cannot explain these morphologies.  
Hydrodynamical simulations of CCSNRs in the framework of the neutrino-driven mechanism do not address point-symmetric morphologies (e.g., \citealt{OrlandoJankaetal2025A, OrlandoJankaetal2025B, OrlandoMicelietal2025, Orlando2026}).  
Therefore, it is utmost important to study the morphologies of CCSNRs and compare them with the predictions of the two theoretical explosion mechanisms. In this study, we conduct JJEM simulations with three pairs of jets to explain the rings and the bar of SNR 11.2-0.3.
In \cite{AkashiSoker2026a}, we reproduced the pair of rings of SNR 11.2-0.3. Here we add one pair of jets to explain the bar of this SNR, i.e., a bright X-ray bar extending across the SNR through its center. We do not aim to explain all features of this CCSNR. In particular, we do not add a pair of jets to explain the second jet axis observed in this CCSNR. 
 
The approach we take is the one that has been successfully used in the study of planetary nebulae and which has lead to a deep understanding of the shaping of a rich variety of jet-shaped planetary nebula morphologies (e.g.,  \citealt{Morris1987, Soker1990AJ, SahaiTrauger1998, AkashiSoker2018,   EstrellaTrujilloetal2019, Tafoyaetal2019, Balicketal2020,   GarciaSeguraetal2020, GarciaSeguraetal2021, Clairmontetal2022, RechyGarciaetal2020, Danehkar2022, MoragaBaezetal2023, Ablimit2024, Derlopaetal2024, Mirandaetal2024, Sahaietal2024, Masaetal2026}), including precessing jets (e.g., \citealt{Guerreroetal1998, Mirandaetal1998, Sahaietal2005, Boffinetal2012, Sowickaetal2017, RechyGarciaetal2019, Guerreoetal2021, Clairmontetal2022}); for a recent review on PNe see \cite{Kwoketal2026Galax}. The first step is an eye inspection and a qualitative classification of planetary nebulae and their morphological features (e.g., \citealt{Balick1987, Parkeretal2006, Sahaietal2007, Kwok2024}). This qualitative approach yields a robust identification of jet-shaped morphological features (e.g., \citealt{SahaiTrauger1998}), which motivate and guide jet-shaped hydrodynamical numerical simulations. The qualitative and partially quantitative comparisons of simulations with observations over the years (e.g., \citealt{Akashietal2018, GarciaSeguraetal2021, GarciaSeguraetal2022, GarciaSeguraetal2025, Akashietal2025, Kastneretal2025} and references to earlier studies therein) have established jets as a major player in powering and shaping planetary nebula outflows. Considering the many morphological similarities between CCSNRs and planetary nebulae (e.g., \citealt{BearSoker2017, Soker2024PNSN, KlimovSoker2026b}), the systematic study of morphological features that 3D hydrodynamical simulations of jets give against those observed in CCSNRs became a major task in studying the JJEM (e.g., \citealt{Braudoetal2025, SokerAkashi2025, AkashiSoker2026a, BraudoSoker2026}). 
Like most hydrodynamical simulations of jet-shaped planetary nebulae, we launch jets by hand, rather than obtain them self-consistently from the accretion inflow. This is a drawback of these types of simulations, as it is extremely difficult to self-consistently obtain jets from an inflow at a large distance from the compact object.  
 
We summarize the main ingredients of our numerical setting in Section \ref{sec:Numerics}; more details are in \cite{AkashiSoker2026a}. In Section \ref{sec:Results} we describe the bar and rings we obtain, and in Section \ref{sec:G11203} we compare our results to the bar and rings of SNR G11.2-0.3.  
We summarize this study in Section \ref{sec:Summary}.

% =========================
\section{The numerical settings}
\label{sec:Numerics}
% =========================

% =========================
\subsection{Initial model}
\label{subsec:Model}
% =========================
Our numerical setting is very similar to that in \cite{AkashiSoker2026a}, where more details can be found; here we describe the basic properties of the numerical setting. The main difference is that we add another pair of wide jets here. In \cite{AkashiSoker2026a}, we launched two consecutive pairs of conical jets into the core and obtained two opposite circum-jet rings. The first pair had wide jets, and the second had narrow jets along a common axis. (We note that \citealt{GarciaSeguraetal2021} simulated alternative narrow and wide jets in post-common envelope binary systems to account for planetary nebula shaping.) The interaction between the two jet pairs and the core material formed two opposing rings. We repeat the same setting, but we add another wide pair of jets, perpendicular to the axis of these two pairs. The specific motivation is to explain the bar in SNR G11.2-0.3, and the general goal is to show the wide variety of observed morphologies that the JJEM can account for. 

The limited numerical resources and the aim to scan the parameter space force us to some simplifications. (1) We launch the jets at thousands of km instead of their origin at tens of km. (2) We launch very energetic jets that accelerate the inner core in a short time to velocities much larger than the escape velocities, and so we neglect gravity.   

We use the Eulerian adaptive-mesh refinement (AMR) code \textsc{FLASH} v4.8 \citep{FryxellEtAl2000} to conduct our 3D hydrodynamical simulations. The computational domain is a Cartesian box $(x,y,z)$ with 
\begin{equation}
    -1.8 \times 10^{10} \cm \le x,y,z \le 1.8 \times 10^{10} \cm,  
\label{eq:size}
\end{equation}
and outflow boundary conditions on the six cube faces.
We apply 7 refinement levels above the base grid, with one additional enforced level in the central region where we inject the jets; this gives a maximum of 8 levels corresponding to an effective resolution of $2^{10}$ cells per dimension and a minimum cell size of $ \Delta x_{\min} = 3.5 \times 10^{7} \cm$. 

Our initial stellar model at $t=0$ is a spherical striped-envelope model adapted from \cite{Braudoetal2025} and \cite{Braudoetal2026}, which we place at the center of the grid.  
The innermost region of the stellar model, $0 \le r \le  2 \times 10^{9}\cm$, is as in \citet{PapishSoker2014Planar}, who fit a $t \simeq 0.2 \s$ post-bounce structure of a $15 M_\odot$ progenitor \citep{Liebendorferetal2005}, with the density $\rho(2\times10^{9}\cm) \simeq 6\times10^{4} \g \cm^{-3}$. In the volume from this radius to the stellar surface, $2\times10^{9}\cm \le r \le 8\times10^{9} \cm$,  the structure is a hydrogen- and helium-stripped envelope of a $15 M_\odot$ Wolf–Rayet progenitor obtained from a MESA simulation. The density decreases to $\simeq 30 \g \cm^{-3}$ near the stellar surface. 
Outside the star, $r > 8\times10^{9} \cm$, there is a circumstellar medium with a decreasing density.  
To prevent numerical difficulties, like too short time-steps, we impose an inner inert core, e.g, we set the velocity to be zero there, in the volume 
\begin{equation}
   r<  R_{\rm core} = 4 \times 10^{8} \cm. 
\label{eq:Rcore}
\end{equation}

% =========================
\subsection{The jittering-jets that explode the star}
\label{subsec:Jets}
% =========================

In this study, we launch three pairs of jittering jets. The substantial difference from our previous study \citep{AkashiSoker2026a} is the addition of the second pair. The motivation came from a careful inspection of the morphology of SNR G11.2-0.3 that we re-examine in Section \ref{sec:G11203}.  Table \ref{Table1} summarizes the properties of the six jets. We launch all jets at an initial velocity of $v_{\rm j} =  80,000 \km \s^{-1}$ at a very high Mach number (the initial thermal energy of the jets is negligible). We inject the jets in a radius of $r_{\rm inj} = 7 \times 10^{8} \cm$.
The first pair of very wide, opposite jets is the most energetic and initiates the explosion. The second pair is another energetic wide pair, perpendicular to the first pair. The third pair consists of narrow jets that interact with the shell formed by the two pairs of jets as they accelerate and compress the core material. We launch the third pair along the same symmetry axis as the first pair to simplify the analysis. However, as we showed in \cite{AkashiSoker2026a}, the narrow pair can be inclined to the first pair, as long as the narrow jets follow some segment of the wide jets.  
% TTTTTTTTTTTTTTTTTTTTTTTTTTTTTTTTTTTTTTTTTTTTTTTTTTTTTTTTTT
\begin{table*}[t]
\begin{center}
\caption{Jet properties}
\begin{tabular}{|l|c|c|cc|c|}
\hline
Jets' pair     & Wa   & Wb        & N-up & N-down& Total \\
\hline
 t(start) [s]  & 0    & 0.5        & 3  & 3  & 0 \\ 
 t(end) [s]    & 0.5  & 1      & 4  & 4  & 4 \\ 
\hline
Axis $(x,y,z)$  & $(\pm 0.5,0, \pm \sqrt{3}/2)$ & $(0, \pm 1 , 0)$  & $(+0.5,0, +\sqrt{3}/2)$  & $(-0.5,0, -\sqrt{3}/2)$ & \\
\hline
$\alpha$   & $75^\circ$  & $40^\circ$  & $10^\circ$  & $5^\circ$  &  \\ 
\hline
$M_{\rm 1j}$ [$M_\odot$]   & 0.024  & 0.0048  & 0.005  & 0.005 & \\
\hline
$E_{\rm kin, 1j}$ [erg]   & $1.5\times10^{51}$ & $0.3\times10^{51}$  &  $0.32\times10^{51}$ & $0.32\times10^{51}$ & $4.24 \times 10^{51} \erg$\\
\hline 
\label{Table1}
\end{tabular}
\end{center}
\begin{flushleft}
\small 
Notes: We launch all jets with an initial velocity of $v_j = 8 \times 10^4 \km \s^{-1}$. The two opposite jets in each of the first two pairs are equal in mass, energy, and half-opening angle. 
Variables: t(start) and t(end) are the starting and end time of the pair of jets; Axis is the symmetry axis of the jet; $\alpha$: half opening angle; $M_{\rm 1j}$: mass in one jet; $E_{\rm kin, 1j}$: Kinetic energy of one jet. The total explosion energy is $E_{\rm exp}=4.24 \times 10^{51} \erg$.     
\end{flushleft}
\end{table*}
% TTTTTTTTTTTTTTTTTTTTTTTTTTTTTTTTTTTTTTTTTTTTT

The wide jets that we launch here can represent wide jets, narrower jets that precess around the symmetry axis, or, more likely, the jittering of narrower jets around an axis. Namely, in one jet-launching episode, a pair of opposite narrower jets jitter around an axis; the effect is like a wide pair of jets. Another effect that can increase the opening angle that a jet induces is the boosting by neutrino heating \citep{Soker2022nu}. In the JJEM, neutrino heating does play a role, but only as a secondary effect in increasing the explosion energy somewhat. After the jets penetrate the stalled shock region, neutrino heating can accelerate more material around the jet, widening the opening angle \citep{Soker2022nu}.  

We comment on the launching of the jets from a distance $\simeq 100$ times as large as their real launching radius (the radius of the proto-Ns), and the omission of gravity. We expect these two strong approximations to have a little effect on the general conclusion because we are not examining fine-detailed structure. The study of smaller structures requires launching jets closer to the center and accounting for gravity, as in the studies of clumps \citep{Braudoetal2025} and pipe morphology \citep{BraudoSoker2026}. The first two pairs of jets set off the explosion. The important outcome for this study is that they explode the core and form a dense shell, and that they accelerate the material less efficiently perpendicular to the plane they define. Since the jets we simulate are much more energetic than a typical CCSN explosion, gravity will have little effect. As well, these energetic jets will easily move from the real launching radius of $50 \km$ to $\simeq 7000 \km$ because most of the original core material from this volume is already in the NS.  
For the shaping of rings by the third pair of jets, the expanding shell is already at $> 10^4 \km$ in reality, and so launching the third pair from $7,000 \km$ instead of $\simeq 50 \km $ has little effect on the results.

% =========================
\section{Results}
\label{sec:Results}
% =========================

In this study, we focus on a specific morphology and will present the results at the end of the simulation, $t=6.3 \s$. In \cite{AkashiSoker2026a}, we simulated cases with two pairs of jets, similar to the first and third in this study, and presented the time evolution, vorticity maps, and Rayleigh-Taylor instability maps for several sets of two-pairs simulations. The basic evolutionary and physical properties of this type of simulation are in that paper. Here, we present one simulation with three pairs of jets, aiming to qualitatively reproduce a morphology that we compare with that of SNR G11.2-0.3 in Section \ref{sec:G11203}. Table \ref{Table1} lists the jets' properties.      

\begin{figure}
\centering
\includegraphics[width=\linewidth, trim=4.5cm 0.0cm 4.5cm 0cm, clip]{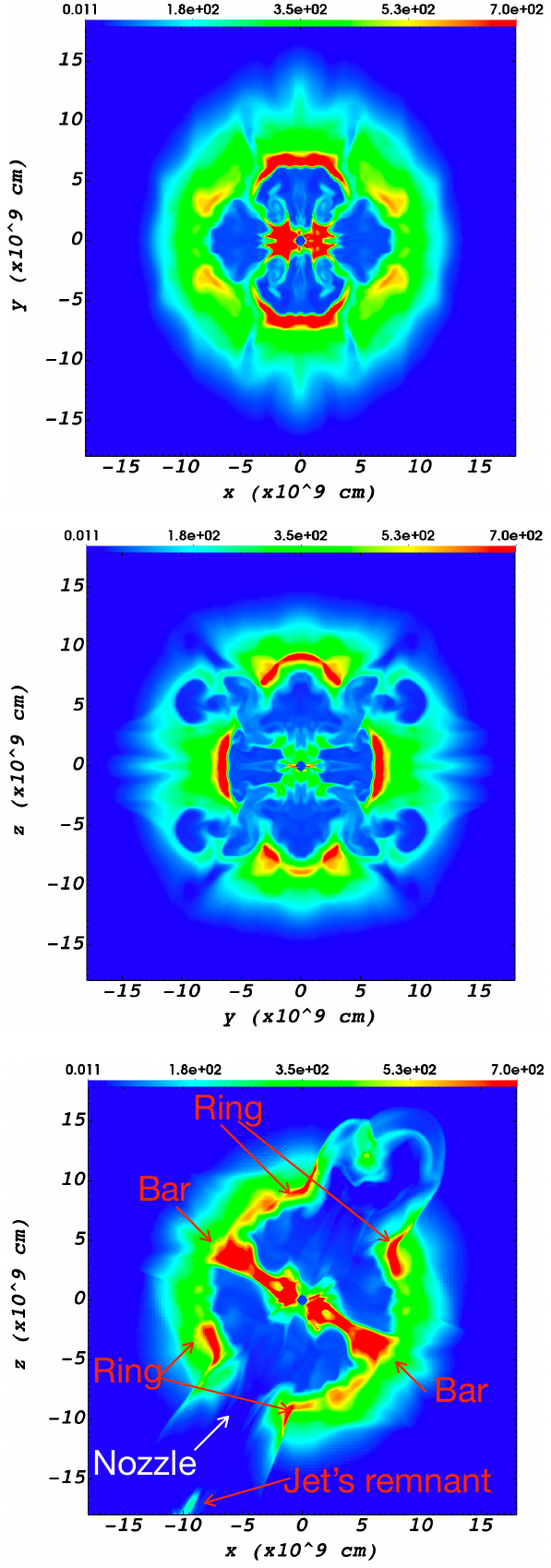}
\caption{Density maps in three Cartesian planes at $t = 6.3~\mathrm{s}$: $z=0$, $x=0$, and $y=0$, from top to bottom. The density is given in units of $\mathrm{g\,cm^{-3}}$ in a logarithmic scale according to the color bar.}
\label{fig:density_maps}
\end{figure}

\begin{figure}
\centering
\includegraphics[width=\linewidth, trim=0.0cm 0.0cm 0.0cm 0.0cm, clip]{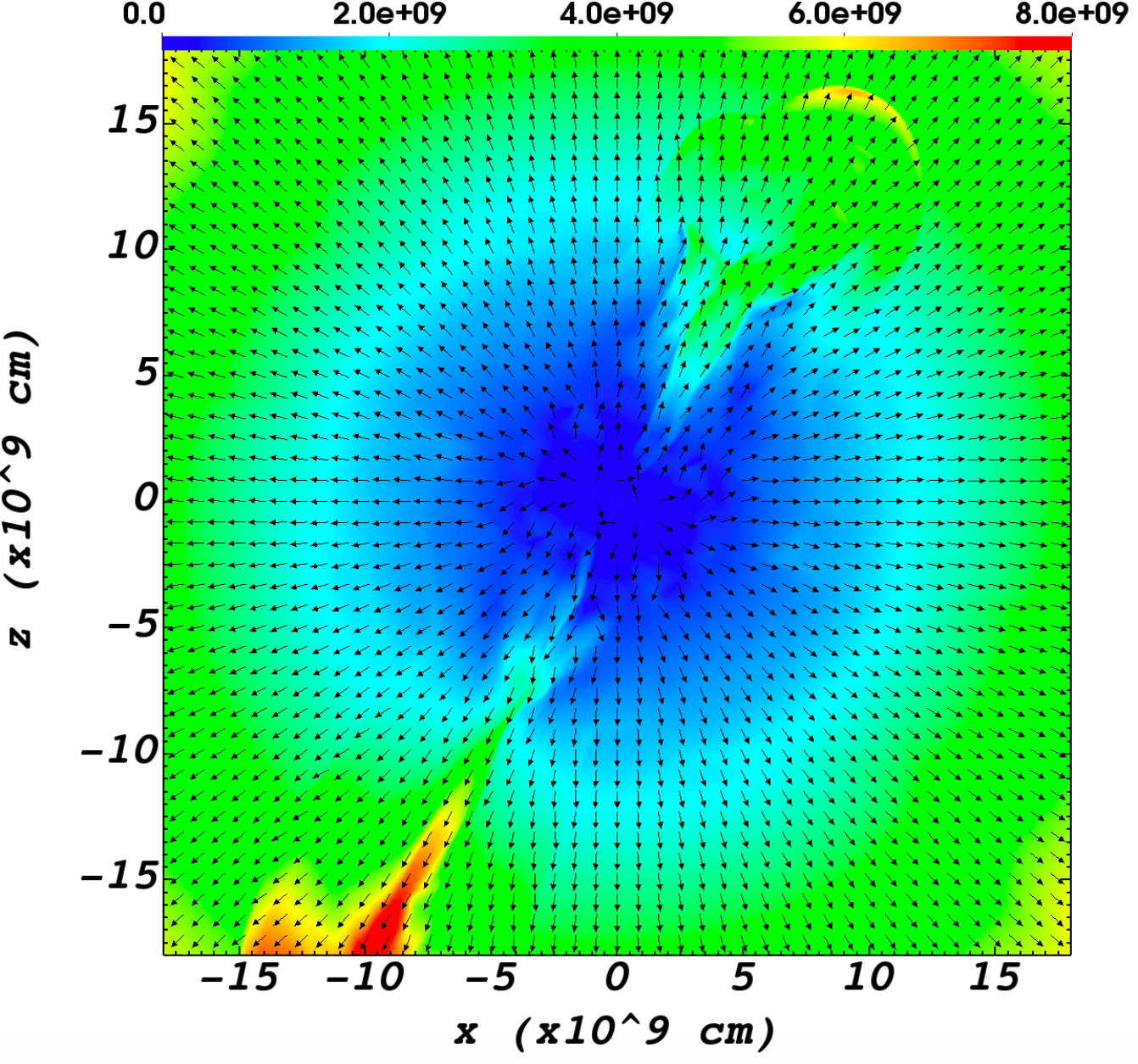}
\caption{Velocity field at $t = 6.3~\mathrm{s}$ in the $y=0$ plane, corresponding to the density map in the lower panel of Figure \ref{fig:density_maps}. Colors depict the velocity magnitude according to the color bar in units of $\mathrm{cm\,s^{-1}}$, and arrows represent the local flow direction.}
\label{fig:velocity_maps}
\end{figure}

Figure \ref{fig:density_maps} presents the density maps in the three grid planes through the center: $z=0$, $x=0$, and $y=0$. The lower panel shows the density map in the $y=0$ plane, which contains the axis of the first pair of jets (wide jets) and the third (narrow). We mark the relevant morphological features on that plane. The first wide pair of jets compresses the core material to form an expanding dense shell (see \citealt{AkashiSoker2026a}). The narrow jets of the third pair interact with this shell, penetrate it, and compress gas to the sides, forming two opposite rings. The two opposite narrow jets have equal energy, but the down jet is narrower, hence it penetrates more and forms a nozzle. This asymmetry is also evident from the velocity map in the plane $y=0$ that we present in Figure \ref{fig:velocity_maps}. The narrow jet maintains its launched velocity as it penetrates, while the up jet is slowed down.     
% FFFFFFFFFFFFFFFFFFFFFFFFFFFFFFFFFFFFFFFFFFFFFFFFFFFFFFFFFFFFFFFFFF
% 
% FFFFFFFFFFFFFFFFFFFFFFFFFFFFFFFFFFFFFFFFFFFFFFFFFFFFFFFFFFFFFFFFFF
% FFFFFFFFFFFFFFFFFFFFFFFFFFFFFFFFFFFFFFFFFFFFFFFFFFFFFFFFFFFFFFFFFF
% FFFFFFFFFFFFFFFFFFFFFFFFFFFFFFFFFFFFFFFFFFFFFFFFFFFFFFFFFFFFFFFFFF

Each of the wide jets of the first pair inflates a bubble. These bubbles compress the core material in the plane between them. This forms a dense equatorial expanding disk \citep{AkashiSoker2026a}. The novel qualitative feature of this study is a second pair of wide jets along the $y$-axis, i.e., perpendicular to the axes of the first and third pairs. This pair accelerates some of the dense disk material from regions outside the $xz$-plane. This forms a bar in the $xz$-plane, as we mark on the lower panel of Figure \ref{fig:density_maps}. The other two panels in Figure \ref{fig:density_maps} do not show the extended dense material.   

\begin{figure}
\centering
\includegraphics[trim=0.0cm 7.8cm 0.0cm 0.0cm ,clip, scale=0.40]{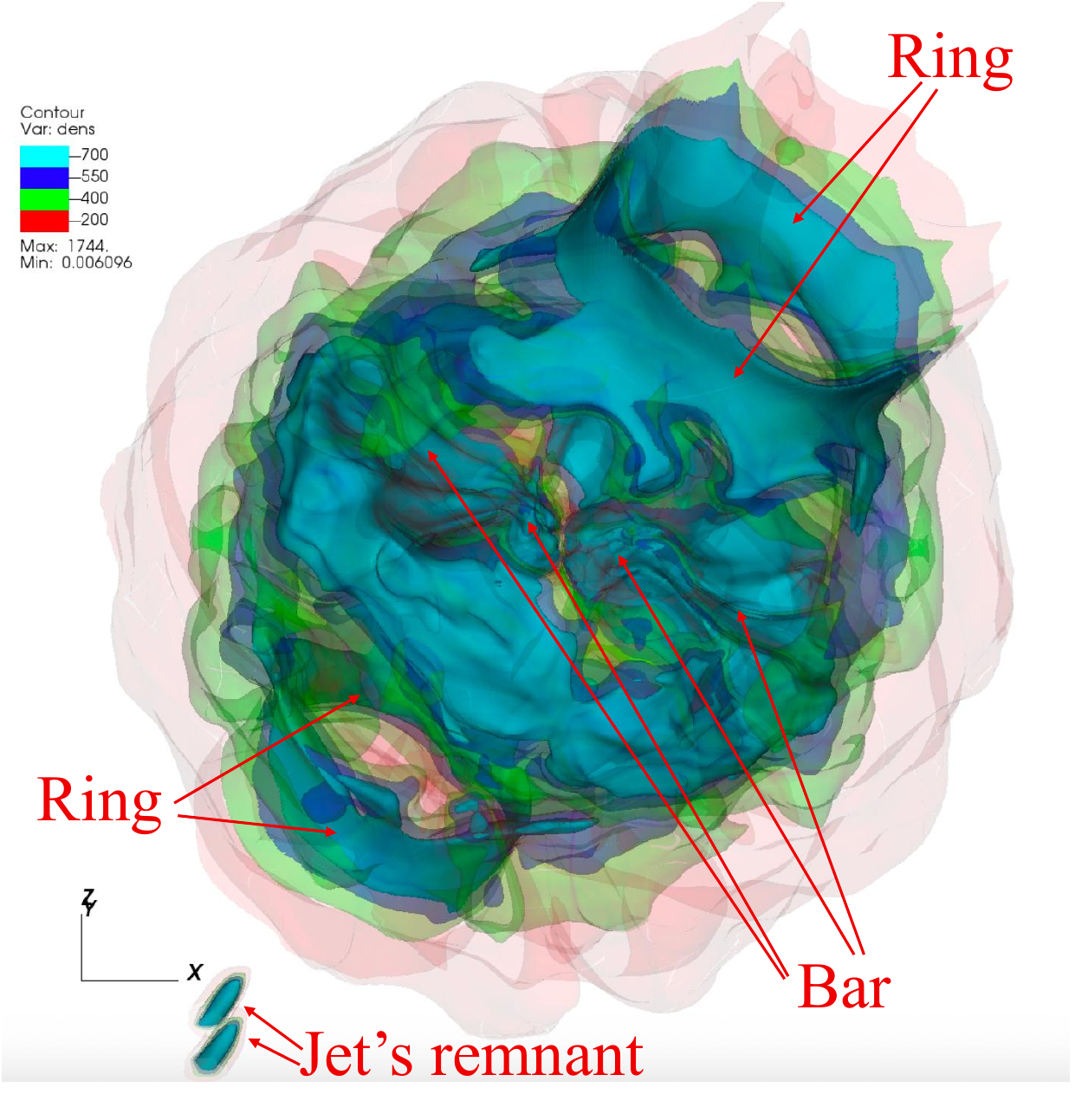} 
\caption{Three-dimensional visualization of the gas density for the simulation at $t=6.3 \s$, composed of four equidensity semi-transparent surfaces as the color bar shows in units of $\g \cm^{-3}$. The viewing direction is at $50^\circ$ to the $z$-axis in the $yz$-plane. We point out the relevant morphological features of this study: two opposite circum-jet rings and the remnant of one jet that shaped the circum-jet down ring, and two opposite dense `tongues' extending from the center outwards; these two dense `tongues' form the bar.  }
\label{fig:3D}
\end{figure}

The main new result of this simulation with respect to those in \cite{AkashiSoker2026a} is the formation of the bar. To further illustrate this structure, we present a 3D visualization of the density in Figure \ref{fig:3D}, composed of four equidensity surfaces. We point at the relevant morphological features. The bar is not straight. This results from the interaction of the three jets and the limited numerical resolution. The axes of the first and third pairs are $30^\circ$ degrees to the $z$-axis in the $y=0$ plane. Namely, not along any symmetry axis of the grid. This can cause asymmetrical numerical effects. The finite resolution causes deviation from complete symmetry around the jets' axis, which is amplified as the gas expands. 
%FFFFFFFFFFFFFFFFFFFFFFFFFFFFFFFFFFF
%FFFFFFFFFFFFFFFFFFFFFFFFFFFFFFFFFFF

\begin{figure}
\centering
\includegraphics[width=\linewidth, trim=2.8cm 2.0cm 3.cm 0cm, clip]{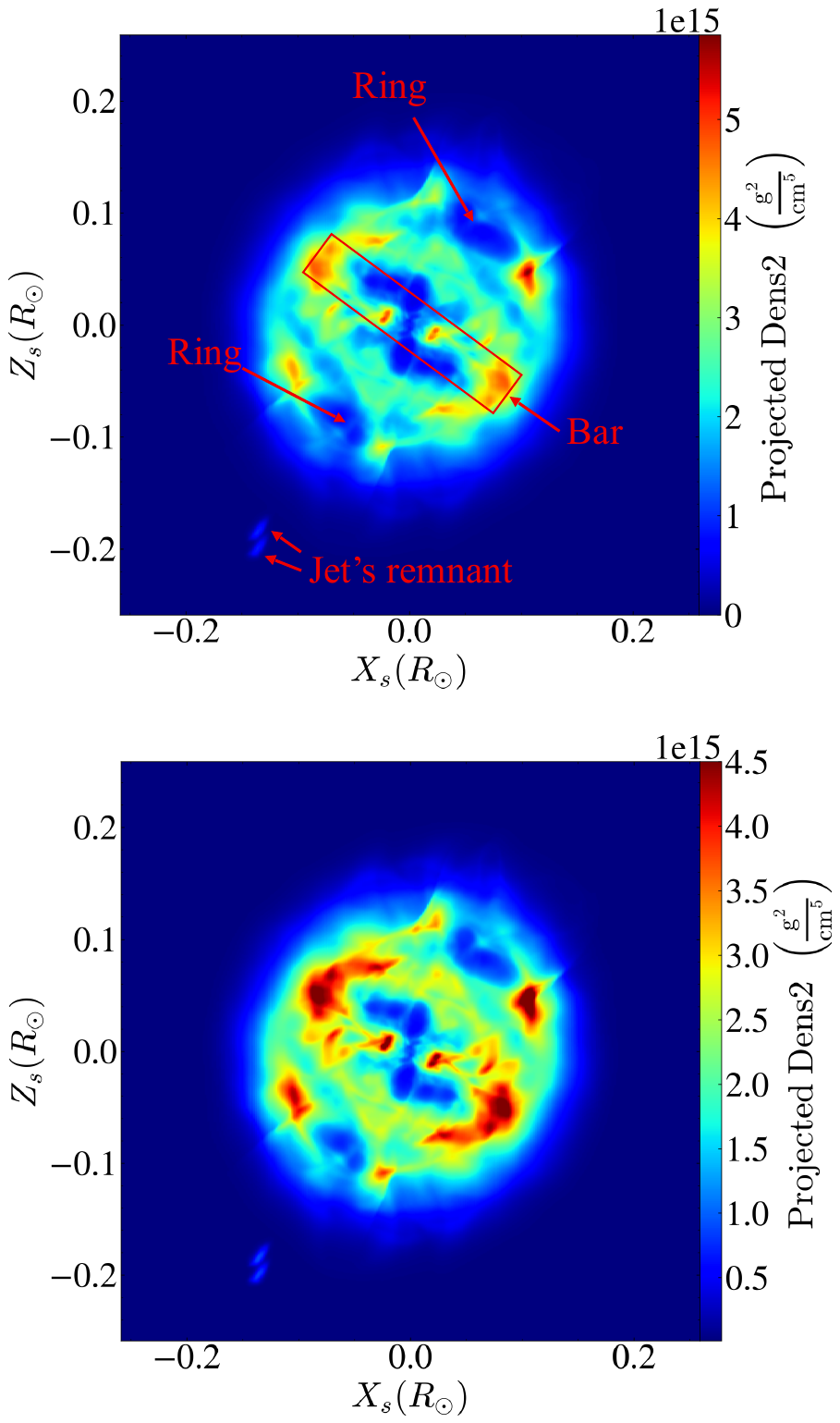}
\caption{Emission measure maps (equation \ref{eq:EI}) for a line of sight in the $zy$ plane and inclined by $50^\circ$ to the $z$-axis. The top panel shows the full-scale distribution, whereas the bottom panel caps the maximum at ${\rm EI} = 4.5 \times 10^{15} \g^2 \cm^{-5}$ to enhance contrast in lower-density regions. The emission measure is expressed in units of $\mathrm{g^2\,cm^{-5}}$.}
\label{fig:proj_50}
\end{figure}

To facilitate comparison with observations, we present in Figure \ref{fig:proj_50} the numerical emission integral   
\begin{equation}
{\rm EI}(X_s,Z_s) \equiv \int \rho^2 dl = \int \rho^2 dY_s,
\label{eq:EI}
\end{equation}
where $\rho$ is the density, and $dl$ is an element along the line of sight; here we take the $Y_s$ to be line-of-sight coordinate, $d Y_s = dl$, such that $X_s Z_s$ is the plane of the sky. The observer is in the $zy$ plane and at $50^\circ$ to the $z$-axis, such that the $x$ and $X_s$ coordinates coincide in these images. The two panels present the same image in different linear scaling. 
In the upper panel, we marked the rings and the bar.
%FFFFFFFFFFFFFFFFFFFFFFFFFFFFFFFFFFF
%FFFFFFFFFFFFFFFFFFFFFFFFFFFFFFFFFFF

There is a large parameter space for which the bar and rings are prominent. We note that for both narrow-angle half-opening angles of $5^\circ$ and $10^\circ$, there are rings. We find that the rings and bar are prominent in these simulations at inclination angles to the $z$-axis of $40^\circ$ to $60^\circ$. We conducted additional simulations involving three pairs of jets with different parameter sets, and in all cases we still obtained the formation of the bar and rings. One representative example is a simulation in which the second pair of wide jets had a half-opening angle of $25^\circ$ instead of $40^\circ$.

% =============================================
\section{Comparison with SNR G11.2-0.3}
\label{sec:G11203}
% =============================================

\begin{figure}[]
	\begin{center}
%	\hspace*{-2cm} 
	% This cut edges: [trim=left bottom right top, clip]{file}
%	\hspace{1cm}
\includegraphics[trim=2.0cm 14.5cm 0.0cm 1.6cm ,clip, scale=0.66]{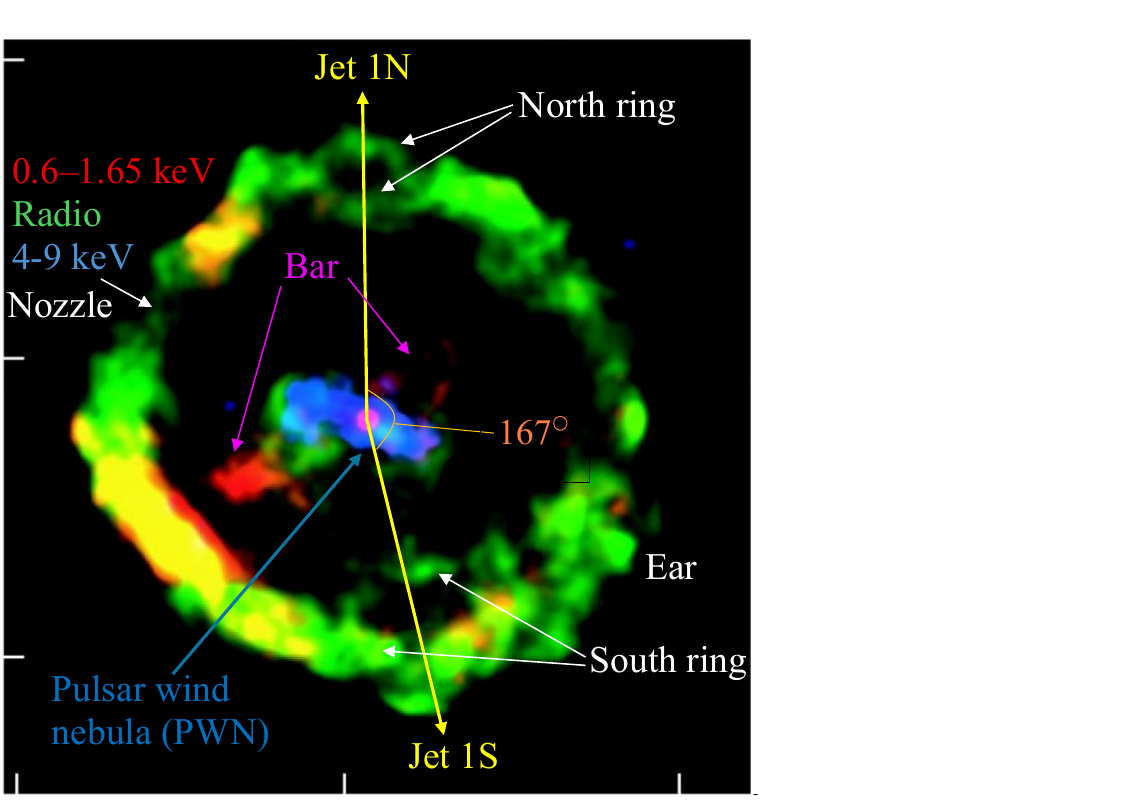} 
\caption{
A figure of SNR G11.2-0.3 adapted from \cite{Robertsetal2003} with mark from \cite{Soker2025G11}, comparing X-ray pulsar wind nebula (PWN) emission and radio emission. Red: $0.6-1.65 \kev$ X-ray. Green: 3.5 cm radio. Blue: $4-9 \keV$ X-ray (all are at $5^{\prime \prime}$ resolution). We added the bar's mark. \cite{Soker2025G11} attributed the shaping of the two rings to a pair of jets, as indicated. For an inclination of the jets' axis to the line of sight of $50^\circ$ \citep{Soker2025G11}, the three-dimensional angle between Jet 1S and Jet 1N is $170 ^\circ$.  
Right ascension (J2000) ticks are 18:11:40, 18:11:30 and 18:11:20, and declination (J2000) ticks are $-19:27:00$, $-19:25:00$, and $-19:23:00$. 
}
%\vskip+0.5cm
\label{Fig:SimG11FigRings1}
\end{center}
\end{figure}

Figure \ref{Fig:SimG11FigRings1} presents an X-ray (red and blue), and radio (green) emission image of SNR G11.2-0.3, adapted from \cite{Robertsetal2003}. The marks for morphological features are from \cite{Soker2025G11}, while we added the bar mark. In \cite{AkashiSoker2026a}, we showed that the narrow jets that we simulated can shape the pair of rings, as we replicated here: Simulated jet N-up corresponds to Jet 1N and N-down to Jet 1S, as \cite{Soker2025G11} marked on the figure.    
% FFFFFFFFFFFFFFFFFFFFFFFFFFFFFFFFFFFFFFFFFF
% FFFFFFFFFFFFFFFFFFFFFFFFFFFFFFFFFFFFFFFFFFF

Although SNR G11.2-0.3 age is $\simeq 1400 - 2400 \yr$ (\citealt{Borkowskietal2016}) and we stop the simulation at $t=6.3 \s$, we can compare our simulation to this SNR because at the and of the simulation ($t=6.3 \s$) the ratio of the kinetic to thermal energy of the ejecta is $E_{\rm kin,ej}/E_{\rm th.ej} = 7.8$. This large ratio implies that the ejecta expansion is already almost homologous, and will keep its structure to very late times (until it accumulates a much larger ISM mass).

The two jets that \cite{Soker2025G11} marked on the observed image in Figure \ref{Fig:SimG11FigRings1} (radial yellow arrows) are bent by $13^\circ$ from being exactly opposite on the plane of the sky. The 3D angle is $170^\circ$ because the 1S-1N jet pair is tilted to the line of sight by $50^\circ$ \citep{Soker2025G11}. Such a bent asymmetry is observed in other CCSNRs, e.g., one pair of bays in the point-symmetric CCSNR the Crab Nebula \citep{ShishkinSoker2025Crab}. The southern ring is larger than the northern one. Asymmetry between two opposite structural features is very common in CCSNRs, because in many cases the two jets of a short-jet-launching episode are expected to be unequal (e.g., \citealt{Soker2024CounterJet}). We argue that the jets that shape the circum-jet rings are jets that participate in the explosion process.  \cite{Soker2025G11} present more analysis of the rings, and in \cite{AkashiSoker2026a} we compared 3D simulations to the observations. Here, we note that the two rings obtained from our new simulation, shown in Figures \ref{fig:density_maps}, \ref{fig:3D}, and \ref{fig:proj_50}, reproduce the general structure and size of the observed rings. The simulated rings are somewhat larger than the observed ones. Narrower jets will form smaller rings, but we cannot simulate such jets due to the limited numerical resolution.
The simulated up-ring is $1.2$ as wide as the down-ring. Crudely, the size goes as the opening angle to the power of $1/3$: $\alpha^{0.3}$. The down-ring diameter is $0.3$ times the ejecta diameter in the simulation. In SNR G11.2-0.3, the north and south rings are $0.19$ and $0.3$ times the shell diameter, respectively. We estimate that a jet with a half-opening angle of $\alpha \simeq 5^\circ$ could have shaped the south ring, and a jet with a half-opening angle of $\alpha \simeq 1^\circ -2^\circ$ could have shaped the north ring. An alternative to such a narrow jet is a cylindrical jet. Namely, the interaction of the jets with the inner core material collimated it to a cylindrical jet. The simulation of this type of interaction requires launching the jets at smaller radii and having a higher resolution. In any case, the range of $\alpha \simeq 1^\circ - 10^\circ$ is as expected in the JJEM \citep{Soker2025Learning}.

We attribute the rings to late jets that participated in the explosion process rather than to post-explosion jets.  
We note that \cite{Gasealahweetal2025} attributed the rings of SNR Circinus X-1 they identified to post-explosion jets rather than exploding jets; in \cite{SokerAkashi2025}, we also argued that explosion-process jets shaped the rings in Circinus X-1.

\begin{figure}[]
	\begin{center}
%	\hspace*{-2cm} 
	% This cut edges: [trim=left bottom right top, clip]{file}
%	\hspace{1cm}
\includegraphics[trim=0.2cm 17.3cm 0.0cm 2.4cm ,clip, scale=0.69]{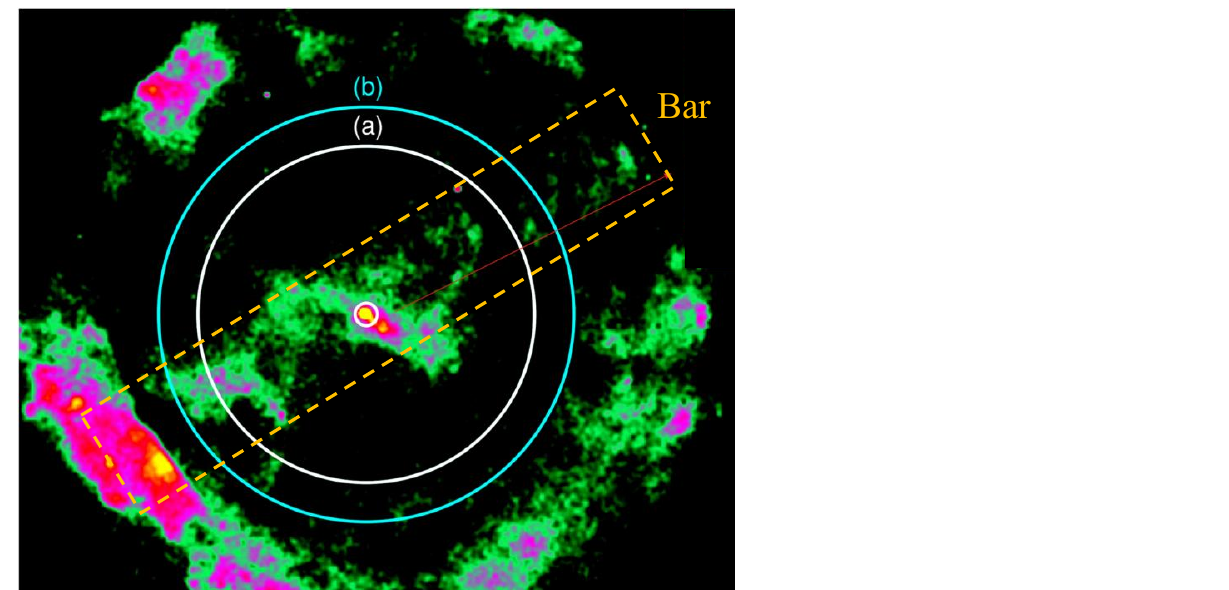} 
\caption{
A Chandra image of SNR G11.2-0.3 in the energy range $0.1-12 \keV$ adapted from \cite{ZhengJTetal2023RAA}, with their marks of the two circles and the red arrow that are irrelevant to this study. The `bar' is the bright inner strip extending from southeast to northwest; we mark its approximate boundary with a dashed orange rectangle. We suggest that two (or more) pairs of wide jets at large angles to the bar axis compressed this material. The bright central point is the pulsar, while the east-west bright rectangle close to the center is the pulsar wind nebula (PWN), seen in blue (hard X-ray) in Figures \ref{Fig:SimG11FigRings1}. }
%\vskip+0.5cm
\label{Fig:SimG11FigureBar1}
\end{center}
\end{figure}

We substantially differ from \cite{Soker2025G11} in the interpretation of the bar of SNR G11.2-0.3. \cite{Soker2025G11} attributed the par to a pair of jets along the axis of the bar, but noted that an accurate axis direction could not be determined. Motivated by our 3D JJEM simulations, we here claim that the bar is not a jet axis, but rather material compressed by two (or more) wide pairs of jets. In Figure \ref{Fig:SimG11FigureBar1}, we mark the approximate boundary of the bar with a dashed-orange rectangle on a Chandra X-ray image from \cite{ZhengJTetal2023RAA}. 
The bar is prominent in Mg K$\alpha$ emission (figure 10 in \citealt{Borkowskietal2016}), suggesting it is Mg-rich and hence was expelled from the inner core, indicating shaping by the explosion process. Our simulations reproduce the bar's clumpy appearance. 
% FFFFFFFFFFFFFFFFFFFFFFFFFFFFFFFFFFFFFFFFFF
% FFFFFFFFFFFFFFFFFFFFFFFFFFFFFFFFFFFFFFFFFFF

\begin{figure}[]
	\begin{center}
%	\hspace*{-2cm} 
	% This cut edges: [trim=left bottom right top, clip]{file}
%	\hspace{1cm}
\includegraphics[trim=0.0cm 15.5cm 0.0cm 0.0cm ,clip, scale=0.56]{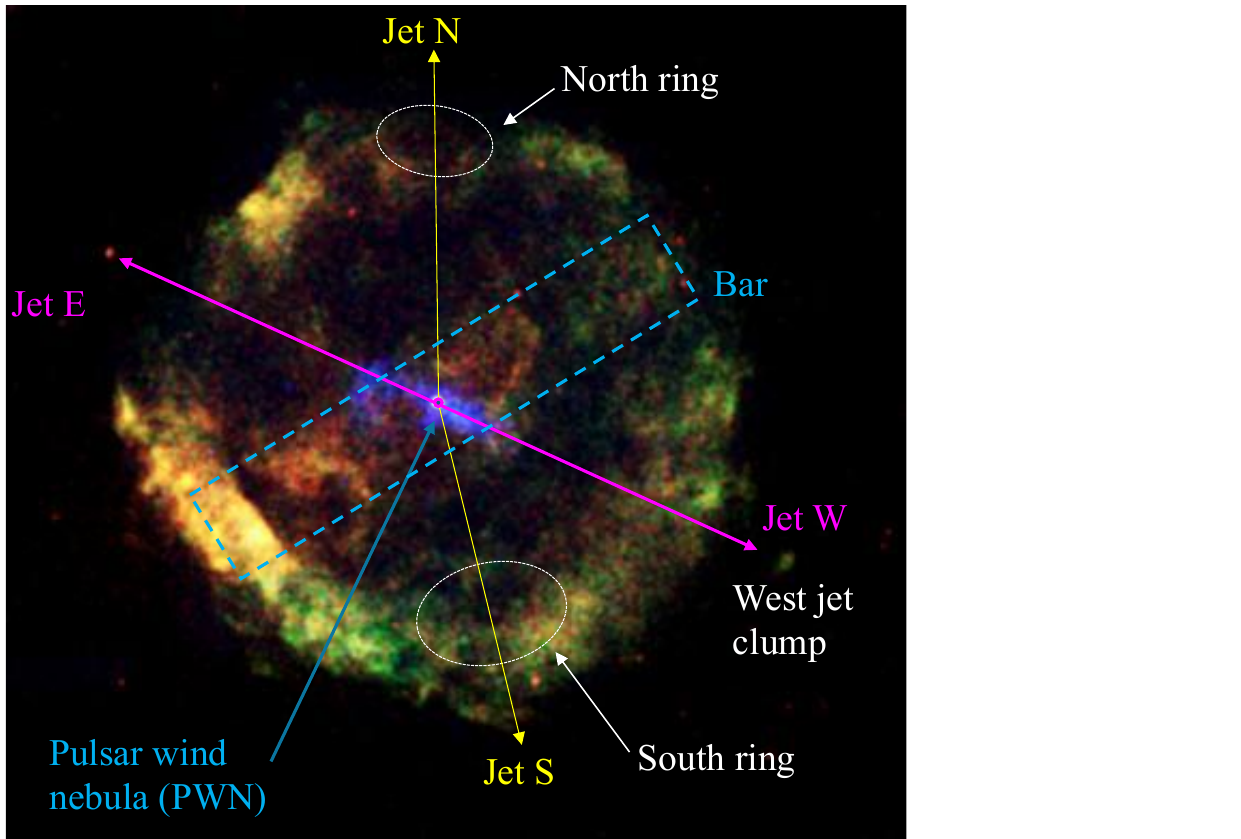} 
\caption{An image of SNR G11.2-0.3 with the two rings that \cite{Soker2025G11} identified and two pairs of jets he assumed shaped the SNR during the explosion. The bar that we identify here replaces a third pair of jets that \cite{Soker2025G11} assumed. The east-west double-sided arrow indicates a pair of jets that shaped the ear-nozzle structure (see Figure \ref{Fig:SimG11FigRings1}) during the explosion (its length is meaningless). The direction of the east-west pair is as the present pulsar jets, but the JJEM assumption is that the jets were active for only a second or so during the explosion. 
We adapted the bare image from the \href{https://chandra.harvard.edu/photo/2007/g11/}{Chandra site}. Colors represent three energy bands: red $0.5-1.5 \keV$; green $1.5-2.5 \keV$; blue $2.5-8 \keV$ 
(Credit: NASA/CXC/Eureka Scientific/\cite{Robertsetal2003}). 
}
%\vskip+0.5cm
\label{Fig:SimG11FigureAll}
\end{center}
\end{figure}

In Figure \ref{Fig:SimG11FigureAll}, we present an X-ray image with the marks of two pairs of jets and the bar. We will not simulate the east-west pair of jets that coincides with the present jet axis of the pulsar. \cite{Soker2025G11} discussed the relation between the east-west pair of jets that participated in the explosion process and the present pulsar jet pair. The structure of the rings, which we attribute to the north-south pair of jets, exists in the X-ray maps (e.g., figure 11 of \citealt{Borkowskietal2016}, and Figure \ref{Fig:SimG11FigureAll}). However, the ring structure is less prominent in the X-ray than in the radio. Differences in appearance across emission bands are prevalent in SNRs and planetary nebulae. For example, radio emission in SNRs depends on magnetic fields, which do not always coincide with the high-density or high-temperature regions that exhibit strong X-ray emission. 
% FFFFFFFFFFFFFFFFFFFFFFFFFFFFFFFFFFFFFFFFFF
% FFFFFFFFFFFFFFFFFFFFFFFFFFFFFFFFFFFFFFFFFFF

% ==================================
\section{Summary} 
\label{sec:Summary}
% ==================================

We conducted simulations of jets exploding the core of an envelope-striped stellar model in the framework of the JJEM. In all simulations, there were three pairs of jets (Section \ref{sec:Numerics}); we present one case (Table \ref{Table1} gives the jets' properties). The first wide pair of jets is very energetic, and it sets off the explosion. The first pair inflates two bubbles that compress material in an expanding shell in most directions, but not perpendicular to the jets' axis. In that plane, the bubbles compress the gas in a plane. The second pair removes material from this plane, beside along a bar that is on an axis perpendicular to the two pairs' axes. We present this bar in Figure \ref{fig:density_maps}. A pair of wide jets can represent two wide jets as we launch, or a collection of pairs of narrow jets jitter at a similar direction.   

The jets of the third pair, now of narrow jets, penetrate the expanding shell that the first pair compressed (Figure \ref{fig:velocity_maps}), and compress material to their sides to form two opposite rings, as we mark on Figures \ref{fig:density_maps}, \ref{fig:3D}, and \ref{fig:proj_50}. 

In Section \ref{sec:G11203} we presented images of SNR G11.2-0.3. We argue that our simulations qualitatively reproduce the bar and the rings of SNR G11.2-0.3. In this study, we did not reproduce the east-west pair of jets that \cite{Soker2025G11} argued for. This would require launching a fourth pair of jets, which, in turn, would require a separate study. We consider the qualitative similarity to somewhat strengthen the JJEM for SNR G11.2-0.3.

 In this study, we simulated a case where the energetic first two pairs of jets were perpendicular to each other. This left slowly expanding material in the direction perpendicular to the plane they defined; this material formed the bar. The important property is that the jets clean the material in a plane. Three or four pairs of jets in one plane could have the same effect. \cite{PapishSoker2014Plan} argued that consecutive pairs of jets stimulate accretion from directions perpendicular to their axes, i.e., the plane they define, and hence the next pair of jets tends to have its axis in the same plane. They also simulated three pairs of jets in the same plane. They showed that when the angle between the axes of the first two pairs of jets is $40^\circ$, no bar forms, whereas for $70^\circ$, a clear bar forms (their figure 6). Therefore, two pairs of jets need not be exactly perpendicular to each other to form a bar. Moreover, if they induce a third pair to be in the same plane, the bar will be more prominent. Overall, the two energetic perpendicular pairs of jets we simulated also represent other cases that add up to form not-so-rare cases.         

To better reproduce observations, future studies should include more processes, which in turn will demand much higher computer resources, including the gravity of the central neutron star, higher resolution, a larger grid to follow the flow to later times, more pairs of jets, and neutrino heating. Simulations of the neutrino-driven mechanism (some of which we cited in Section \ref{sec:intro}) have shown that neutrino heating is significant. Although in the JJEM jets explode the star and play the primary role in the explosion, neutrino heating can boost the explosion \citep{Soker2022nu}. Neutrino heating does not play a primary role, though. Even when neutrino heating might trigger an explosion, the JJEM will operate earlier, and jets will trigger the explosion rather than the stalled shock being revived by neutrino heating \citep{WangShishkinSoker2025}.  Eventually, simulations should show that the accreted material onto the newly-born NS (proto-NS) can form intermittent accretion disks that launch jittering jets. However, it seems that current numerical simulations are not capable of that because magnetic field reconnection, a crucial process in magnetic field amplification and dissipation, requires a numerical resolution beyond what is presently reachable when studying such inflows \citep{Soker2025Learning}.  

Our study is another small step forward in establishing the JJEM as the primary explosion mechanism of CCSNe. 

% ======================================
\section*{Acknowledgements}
% ======================================

 We thank an anonymous referee for helpful comments that improved and extended the presentation of our results. A grant from the Pazy Foundation 2026 supported this research. 
NS thanks the Charles Wolfson Academic Chair at the Technion for the support.

% =================================
% =================================

% =================================
% =================================
% =================================

%%% Below is for using the bib file

% =================================
% =================================
% =================================

%\bibliographystyle{mnras}

%\bibliography{PASPsample701}{}
%\bibliographystyle{aasjournalv7}

\bibliography{BibReference}{}
\bibliographystyle{aasjournalv7}

\end{document}